\documentclass[12pt]{article}

\def\xma{x_{-,A}}
\def\xms{x_{-,S}}

\def\pp{\mu  }

\begin{document}

\begin{titlepage}
\begin{flushright}
CALT-68-2541\\
ITEP-TH-07/05
\end{flushright}

\begin{center}
{\Large\bf $ $ \\ $ $ \\
Plane wave limit of local conserved charges}\\
\bigskip\bigskip\bigskip
{\large Andrei Mikhailov\footnote{e-mail: andrei@theory.caltech.edu}}
\\
\bigskip\bigskip
{\it California Institute of Technology 452-48,
Pasadena CA 91125 \\
\bigskip
and\\
\bigskip
Institute for Theoretical and 
Experimental Physics, \\
117259, Bol. Cheremushkinskaya, 25, 
Moscow, Russia}\\

\vskip 1cm
\end{center}

\begin{abstract}
We study the plane wave limit of the B\"acklund transformations 
for the classical string in AdS space times a sphere and
obtain an explicit expression for the local conserved charges. 
We show that the Pohlmeyer charges become in the plane wave
limit the local integrals of motion of the free massive field.
This fixes the coefficients in the
expansion of the anomalous dimension as the sum of the Pohlmeyer
charges. 
\end{abstract}

\end{titlepage}

\section{Introduction.}
AdS/CFT correspondence is a strong/weak coupling duality.
It maps the strongly coupled ${\cal N}=4$ Yang-Mills theory 
to the Type IIB superstring theory on $AdS_5\times S^5$. 
It is a useful tool  for studying the dynamics of the
strongly coupled Yang-Mills theory, but it would be even more
important to use this duality to learn more about the string
theory.  AdS/CFT correspondence
relates the string theory to the gauge field theory which at this time is
understood much better then the string theory. It would be useful
to learn how to rewrite the gauge theory in stringy degrees of freedom.

Recently the classical strings in $AdS_5\times S^5$ were identified
on the gauge theory side as coherent states in the sector
of single-trace operators with the large R-charge
\cite{FT02,Tseytlin,Russo,MinahanZarembo,FT03,FTQ,Kruczenski}.
The dynamics of such states is essentially classical because
they are composed of the large number of partons (a large number
of degrees of freedom). 

It was conjectured that the
number of partons is in the classical limit a conserved quantity
\cite{Minahan}. We have argued in \cite{Notes} (using the results
of \cite{ArutyunovStaudacher,Engquist,KT}) that this conserved
quantity corresponds to a hidden symmetry of the classical
string theory in $AdS_5\times S^5$. Classical string in $AdS_5\times S^5$
is an integrable system 
\cite{MandalSuryanarayanaWadia,BPR,KMMZ,KazakovZarembo,BKS,SchaferNameki}. 
An infinite family of local
conserved charges
was found by Pohlmeyer \cite{Pohlmeyer}.
The hidden symmetry corresponding to the length of the operator
is generated by an action variable; it is an infinite linear combination of
the Pohlmeyer charges. 

In this paper we will fix the coefficients of this infinite linear
combination by considering the plane wave limit \cite{BMN}. In the
plane wave limit
the worldsheet theory becomes the theory of free massive fields.
Our action variable in this limit counts the total number
of oscillators\footnote{The oscillator number for strings in AdS
was previously discussed in \cite{deVegaLarsenSanchez}. I want
to thank A.~Tseytlin for pointing my attention to this work.}
in the excited state. We will apply the
B\"acklund transformation in the plane wave limit
and obtain in this limit an explicit expression for
the generating function
of the local conserved charges.
We will find that the Pohlmeyer charges become the local conserved
charges of the free massive fields, and that the oscillator
number is indeed an infinite linear combination of them.

The explicit expression for the action variable and the arguments of 
\cite{Anomalous} allow us to answer the
following question: given the classical string worldsheet
in $AdS_5\times S^5$, how to compute the anomalous dimension
of the corresponding field theory operator?
The answer is given by Eq. (\ref{LogC}).

Integrability is a very useful feature of the AdS/CFT correspondence,
and the plane wave limit provides an interesting tool for 
understanding the integrable structures. The worldsheet sigma-model can be 
understood in some regime as an integrable perturbation of the free massive 
theory. Integrable structure of the worldsheet theory was
studied from this point of view in \cite{Alday,SwansonMay,SwansonOct}.
In particular, the plane wave limit of the
commuting conserved charges (and the $1/R^2$ corrections)
was studied on the classical and quantum level in \cite{SwansonMay,SwansonOct}.
The way we use the plane wave limit here is somewhat similar
to considerations in Sections 2 and 3 of \cite{KRT}.

\section{$AdS_5\times S^5$ and the plane wave.}
\subsection{A special choice of coordinates in $AdS_5\times S^5$.}
Let us consider the embeddings of $AdS_5$ as the hyperboloid 
$AdS_5\subset {\bf R}^{2+4}$  
and $S^5$ as  the sphere
$S^5\subset {\bf R}^6$. Let $X_{-1},X_0,\ldots,X_4$
denote the coordinates of ${\bf R}^{2+4}$
and $Y_1,\ldots, Y_6$ the coordinates of ${\bf R}^6$.
The hyperboloid and the sphere are given by the equations		
$$X_{-1}^2+X_0^2-\sum\limits_{i=1}^4 X_i^2 =1,\;\;\;\;\;
\sum\limits_{i=1}^6 Y_i^2=1$$
The metric of $AdS_5\times S^5$ is:
\begin{equation}
	ds^2=R^2\left[-dX_{-1}^2-dX_0^2+\sum\limits_{I=1}^4
	dX_I^2+\sum\limits_{J=1}^6 dY_J^2\right]
\end{equation}
We will consider the string localized near the equator of
the sphere. The following parametrization of $AdS_5\times S^5$
is useful:
\begin{eqnarray}
	&& X_{-1}+iX_0 = 
	\left(1+\epsilon^2 \sum\limits_{j=1}^4 x_j^2\right)^{1\over 2}
	\exp \left[ i\left(x_+ - {\epsilon^2\over 2}x_-\right)
	 \right]\nonumber\\[5pt]
	&& Y_5+iY_6=
	\left( 1-\epsilon^2\sum\limits_{j=1}^4 y_j^2  \right)^{1\over 2}
	\exp \left[ i\left(x_+ + {\epsilon^2\over 2}x_-\right)
	\right]\label{Lift}
	\\[5pt]
	&& X_i=\epsilon x_i,\;\;\; Y_i=\epsilon y_i,\;\;\;(i=1,2,3,4)\nonumber
\end{eqnarray}
Here $\epsilon$ is a small parameter which is usually
chosen to be of the order $ 1/R$.
We will use the conformal coordinates on the worldsheet.
This means that the embedding 
$x(\tau,\sigma)$ satisfies the constraints:
\begin{equation}\label{Virasoro}
	\begin{array}{l}
(\partial_{\tau}X)^2+(\partial_{\sigma}X)^2+
(\partial_{\tau}Y)^2+(\partial_{\sigma}Y)^2=0\\[5pt]
(\partial_{\tau}X,\partial_{\sigma}X)+
(\partial_{\tau}Y,\partial_{\sigma}Y)=0
\end{array}
\end{equation}
With these constraints the equations of motion are:
\begin{equation}\label{EqM}
	\begin{array}{l}
(\partial_{\tau}^2-\partial_{\sigma}^2)
	X-\left[ (\partial_{\tau}X,\partial_{\tau}X)-
	(\partial_{\sigma}X,\partial_{\sigma}X)\right]X=0
\\[5pt]
	(\partial_{\tau}^2-\partial_{\sigma}^2)
	Y+\left[ (\partial_{\tau}Y,\partial_{\tau}Y)-
	(\partial_{\sigma}Y,\partial_{\sigma}Y)\right]Y=0
	\end{array}
\end{equation}
In the plane wave perturbation theory the 
embedding $x(\tau,\sigma)$ has a regular expansion in powers of $\epsilon$.
Let us describe the first approximation.

\subsection{An approximate solution for the classical string.}
Let $\mu$ be a constant parameter.
There is an approximate solution for which the eight ``transverse''
coordinates
 $x_1,\ldots, x_4,y_1,\ldots, y_4$ satisfy
the plane wave equations of motion
\begin{equation}\label{PWEqM}
	\begin{array}{l}
	(\partial^2_{\tau}-\partial^2_{\sigma})x_i+ \mu ^2 x_i=0\nonumber\\[5pt]
	(\partial^2_{\tau}-\partial^2_{\sigma})y_i+ \mu ^2 y_i=0
\end{array}
\end{equation}
with the constraint 
\begin{equation}\label{PeriodicityConstraint}
\int_0^{2\pi}d\sigma 
\left[ (\partial_{\tau}x_i,\partial_{\sigma}x_i)+
(\partial_{\tau}y_i,\partial_{\sigma}y_i)\right]=0
\end{equation}
To describe this solution we have to define 
$x_+(\tau,\sigma)$ and $x_-(\tau,\sigma)$.
Let us introduce the functions $\xma(\tau,\sigma)$ and $\xms(\tau,\sigma)$ 
defined by  the equations:
\begin{eqnarray}
	&&\partial_{\tau}\xma=-{1\over 2 \mu }
	\sum\limits_{i=1}^4\left[(\partial_{\tau}x_i)^2
	+(\partial_{\sigma}x_i)^2-
	\mu ^2 x_i^2 \right]-\partial_{\tau}\phi\\[5pt]
	&&\partial_{\sigma}\xma=
	-{1\over \mu }\sum\limits_{i=1}^4
	(\partial_{\tau}x_i,\partial_{\sigma}x_i)-\partial_{\sigma}\phi
\end{eqnarray}
\begin{eqnarray}
	&&\partial_{\tau}\xms=-{1\over 2 \mu }
	\sum\limits_{i=1}^4\left[(\partial_{\tau}y_i)^2
	+(\partial_{\sigma}y_i)^2-
	\mu ^2 y_i^2 \right]+\partial_{\tau}\phi\\[5pt]
	&&\partial_{\sigma}\xms=
	-{1\over \mu }\sum\limits_{i=1}^4
	(\partial_{\tau}y_i,\partial_{\sigma}y_i)+\partial_{\sigma}\phi
\end{eqnarray}
where $\phi$ is an arbitrary function of $\tau,\sigma$
satisfying 
$(\partial_{\tau}^2-\partial_{\sigma}^2)\phi=0$.
The results of the calculations in this paper do not depend
on the choice of $\phi$. 
We will use $\phi=c\sigma$ with a 
constant\footnote{A nonzero $c$ may be needed if we want $x_{-,A}$ and
$x_{-,S}$ to be periodic functions of $\sigma$. 
The sum $x_{-,S}(\tau,\sigma)+x_{-,A}(\tau,\sigma)$
is periodic in $\sigma$  because of 
(\ref{PeriodicityConstraint}).}
$c$. We put:
\begin{equation}
	\begin{array}{l}
x_+=\mu (\tau-\tau_0)+{\epsilon^2\over 2}(\xms-\xma)\\[5pt]
x_-=(\xms+\xma)
\end{array}
\end{equation}
This solves (\ref{EqM}) approximately, 
the right hand side of (\ref{EqM}) is of
the order $\epsilon^3$. It is possible to modify the
definition of $x_{-,A}$ and $x_{-,S}$ by the terms
of the order $\epsilon^2$, so that the modified
solution solves the constraints (\ref{Virasoro}) exactly.
These corrections will not spoil the equations of
motion (\ref{EqM}) in the order $\epsilon^2$; we will
not need the explicit formula for these correcting terms
in this paper.

\subsection{Relation between $\mu$ and angular momentum.}
The string worldsheet action in the plane wave region is:
\begin{eqnarray}
	S={1\over 2}{1\over 2\pi}\int d\sigma d\tau 
	\left[ 2\partial_{\tau}x_+ \partial_{\tau}x_-
	-(\partial_{\tau}x_+)^2(x^2+y^2)+
	(\partial_{\tau}x)^2+(\partial_{\tau}y)^2+
	\right.
	\nonumber\\[5pt]
	\left. +\epsilon^2\partial_{\tau}x_+ \partial_{\tau}x_- (x^2-y^2)+
	{\epsilon^2}\left((y,\partial_{\tau}y)-(x,\partial_{\tau}x)^2
	\right)+\ldots\right]
	\nonumber
\end{eqnarray}
Here $\ldots$ denotes the terms which do not contain time derivatives
and the terms of the higher order in $\epsilon^2$.
We will write $y^2$ or $(y,y)$ instead of $\sum\limits_{j=1}^4 y_j y_j$.
The vector field $\partial\over\partial x_-$  is a  Killing
vector. The corresponding momentum up to the terms of the order
$o(\epsilon^2)$ is 
\begin{eqnarray}
&&	M\left( {\partial\over\partial x_-} \right)=
	{1\over 2\pi}\int d\sigma\;
	\partial_{\tau}x_+\left(1+{\epsilon^2\over 2}(x^2-y^2)\right)
	=\nonumber\\[5pt]
&&	=\mu-{\epsilon^2\over 4\mu}{1\over 2\pi}\int d\sigma
	\left[ (\partial_{\tau}y)^2+(\partial_{\sigma}y)^2+\mu^2 y^2
	-(\partial_{\tau}x)^2-(\partial_{\sigma}x)^2-\mu^2 x^2\right]
	\nonumber
\end{eqnarray}
The Killing vector field 
${1\over \epsilon^2}{\partial\over\partial x_-}+
 {1\over 2}{\partial\over\partial x_+}$ is periodic with the
 period $2\pi$. Let us denote $J$ the corresponding angular
 momentum.  We have:
 \begin{equation}
	 M\left( {\partial\over\partial x_-} \right)=
	 \epsilon^2
	 \left(J-{1\over 2}M\left({\partial\over\partial x_+}  \right)
	 \right)
 \end{equation}
 The momentum corresponding to 
 ${\partial\over\partial x_+}$ up to the terms of the order $\epsilon^2$ is 
\begin{eqnarray}
&&	 M\left({\partial\over\partial x_+}  \right)=
	{1\over 2\pi}\int d\sigma\left[
	\partial_{\tau}x_- -\partial_{\tau}x_+(x^2+y^2)\right] =\nonumber\\[5pt]
	 &&	 =-{1\over 2\mu}{1\over 2\pi}\int d\sigma 
	 \left[ (\partial_{\tau}y)^2+(\partial_{\sigma}y)^2+\mu^2 y^2
	+(\partial_{\tau}x)^2+(\partial_{\sigma}x)^2+\mu^2 x^2 \right]
	\nonumber
\end{eqnarray}
Therefore
\begin{equation}\label{MuAndJ}
	\mu=\epsilon^2\left( J+{1\over 2\mu}{1\over 2\pi}\int d\sigma 
	 \left[ (\partial_{\tau}y)^2+(\partial_{\sigma}y)^2+\mu^2 y^2
	 \right]\right)+o(\epsilon^2)
 \end{equation}

\section{B\"acklund transformations and Pohlmeyer charges.}
\subsection{B\"acklund transformations.}
We will restrict ourselves to the case when $x_1=\ldots=x_4=0$, which
means that the string fluctuates only in $S^5$; the projection
of the worldsheet to $AdS_5$ is a fixed timelike geodesic.
We will discuss the general case $x_i\neq 0$ at the end of the
paper.
Consider the B\"acklund transformation which does not
change the AdS part of the worldsheet $X$ but changes 
the $S^5$ part $Y\to Y'$:
\begin{eqnarray}
	\partial_+(Y'+Y)={1\over 2}(1+\gamma^{-2})(Y',\partial_+Y)(Y'-Y)
	\nonumber\\[-3pt]\label{BT} \\[-9pt]
	\partial_-(Y'-Y)=-{1\over 2}(1+\gamma^2)(Y',\partial_-Y)(Y'+Y)
	\nonumber
\end{eqnarray}
Here $\gamma$ is a constant parameter. We will also
use the notation $Y(\gamma)=Y'$. Notice that 
\begin{equation}\label{ConstantAngle}
	(Y',Y)={1-\gamma^2\over 1+\gamma^2}
\end{equation}
Let us apply the transformation $Y\to Y'$ in the plane wave limit. 
At the zeroth order 
in $\epsilon^2$ we get:
$$(Y',\partial_+Y)=-{\mu \over 2}\sin[\mu (\tau'_0-\tau_0)]+O(\epsilon^2)$$ 
and $(Y',\partial_-Y)=(Y',\partial_+Y)+O(\epsilon^2)$.
Eqs. (\ref{BT}) at the zeroth order in $\epsilon^2$:
\begin{eqnarray}
&&	\partial_{\tau}
	\left(e^{i\mu (\tau-\tau'_0)}+e^{i\mu (\tau-\tau_0)}\right)=
	-{\mu \over 2}(1+\gamma^{-2})
	\sin[\mu (\tau'_0-\tau_0)]
	\left(e^{i\mu (\tau-\tau'_0)}-e^{i\mu (\tau-\tau_0)}\right)
	\nonumber\\[5pt]
&&	\partial_{\tau}
	\left(e^{i\mu (\tau-\tau'_0)}-e^{i\mu (\tau-\tau_0)}\right)=
	{\mu \over 2}(1+\gamma^2)\sin[\mu (\tau'_0-\tau_0)]
	\left(e^{i\mu (\tau-\tau'_0)}+e^{i\mu (\tau-\tau_0)}\right)
	\nonumber
\end{eqnarray}
This implies that $\tan\left[{\mu \over 2}(\tau'_0-\tau_0)\right]=\pm\gamma$.
We will choose the plus sign\footnote{The relations (\ref{BT}) are
symmetric in $Y$ and $Y'$. Therefore there are two solutions for $Y'$.
Given one solution $Y'=Y(\gamma)$ the other one is $Y(-\gamma)$.}.
Let us introduce $\alpha$: 
$$\tan\alpha=\gamma$$
With this notation $\tau_0'=\tau_0+2\mu^{-1}\alpha$.
Then up to the terms of the order $\epsilon^2$ we have:
\begin{equation}
	y'={\mu -2\tan\alpha\;\partial_+\over \mu +2\tan\alpha\;\partial_+}\;y=
	{2\partial_-+\mu \tan\alpha\over 2\partial_--\mu \tan\alpha}\;y
\end{equation}
We can use (\ref{ConstantAngle}) to find the transformation 
$x_{-,S}\to x_{-,S}'$:
\begin{equation}
	x_{-,S}'-x_{-,S}={\cos 2\alpha\over 2\sin 2\alpha}
	(y^2+(y')^2)-{1\over \sin 2\alpha}(y,y')
\end{equation}
The parameter $\mu^{-1}$ is the small parameter of the null-surface
perturbation theory. In the leading order in $1/\mu$
we have $\partial_+\simeq {1\over 2}\partial_{\tau}$ and
the B\"acklund transformation is just a time shift:
\begin{eqnarray}\label{FirstApproximation}
Y'&=&\exp(-2\alpha\mu^{-1}\partial_{\tau})Y=
	\cos(2\alpha)Y-\sin(2\alpha)\mu^{-1}\partial_{\tau}Y
\\[5pt]
&=&{1-\gamma^2\over 1+\gamma^2}Y-
	{2\gamma\over 1+\gamma^2}
	{\partial_{\tau}Y\over |\partial_{\tau}Y|}
	\nonumber
\end{eqnarray}
This formula follows directly from (\ref{BT}).
We have  used that $\partial^2_{\tau}Y=-\mu^2 Y$ in the limit $\mu\to\infty$.
Therefore we should consider $\mu^{-1}\partial_{\tau}$ to be of the order
$1$ in the $1/\mu$-expansion.
We can define the B\"acklund transformations in 
the null-surface perturbation theory as perturbative
solutions of (\ref{BT}) which are power series
in $\mu^{-1}$, the first approximation given by (\ref{FirstApproximation}).
The corrections to (\ref{FirstApproximation})
by the higher powers of $\mu^{-1}$ involve
higher derivatives in $\tau$ and $\sigma$
and depend on $\gamma$ as rational functions.
When $\gamma$ is small we can expand these corrections 
in powers of $\gamma$, and when $\gamma$ is large in powers
of $\gamma^{-1}$.
Therefore the definition of the B\"acklund transformation
as a power series in $1/\mu$ agrees with the usual 
definition as a power series in $\gamma$ or $1/\gamma$.

\subsection{Pohlmeyer charges.}
The generating function for the conserved charges is given by:
\begin{eqnarray}
	{\cal E}(\gamma)&=\;\;{1\over 2\pi}\int d\sigma& 
	\left[\gamma(Y(\gamma),\partial_+Y)+\gamma^3(Y(\gamma),\partial_-Y)
	\right]= \label{GGF}
\\[5pt]
	&=\; \gamma {1\over 2\pi}\int d\sigma &
	\left\{ -{\sin 2\alpha\over 2\cos^2\alpha}\mu +\right.  \nonumber
\\[5pt]
&&	+\epsilon^2\left[{1\over 2\cos^2\alpha}(y',\partial_{\tau}y)+
	{1\over 2}{\cos 2\alpha\over \cos^2\alpha} (y',\partial_{\sigma}y)
	+\right.\nonumber\\[5pt]
&&	\;\;\;\;\;\;+{1\over 2}{\sin 2\alpha\over \cos^2\alpha}\left[
	{\mu \over 2}(y^2+(y')^2)-\partial_{\tau}x_{-,S}\right]+
	\nonumber\\[5pt]
&&	\left.\left.\;\;\;\;\; 
+{1\over 2}{\cos 2\alpha\over \cos^2\alpha} \left[
\mu (x'_{-,S}-x_{-,S})-(\partial_{\tau}y,y)\right]\right]\right\}
	\label{GeneratingFunction}
\end{eqnarray}
The free field  $y$ has an oscillator expansion:
\begin{equation}
	y^i(\tau,\sigma)=\sum\limits_{n=-\infty}^{\infty}
	\sum\limits_{i=1}^4 {1\over \sqrt{2\omega_n}}
	\left(\alpha_n^i e^{in\sigma + i\omega_n\tau}+
	\overline{\alpha_n^i}e^{-in\sigma-i\omega_n\tau}\right)
\end{equation}
where $\omega_n=\sqrt{\mu^2+n^2}$.
We will suppress the index $i$ and write $\sum_n \alpha_n\overline{\alpha_n}$
instead of $\sum\limits_{n=-\infty}^{\infty}\sum\limits_{i=1}^4
\alpha^i_n\overline{\alpha_n^i}$. 
The B\"acklund transformation for the oscillators is:
\begin{equation}
	y'={\mu -2\tan\alpha\partial_+\over
	\mu +2\tan\alpha\partial_+}y\; \Rightarrow\;
	\alpha'_n={\pp-i\tan\alpha (\omega_n+n)\over 
	\pp+i\tan\alpha(\omega_n+n)}\alpha_n
\end{equation}
The oscillator expansion of (\ref{GeneratingFunction}) is:
\begin{equation}
	{\cal E}(\gamma)=-\gamma^2\mu + 
	\epsilon^2{\gamma^2\over \mu }{\cal H}_2+\epsilon^2\gamma^2
	\hat{\cal E}_+(\gamma)
\end{equation}
where 
\begin{equation}
	{\cal H}_2={1\over 4\pi}\int d\sigma \left[ (\partial_{\tau}y)^2+
	(\partial_{\sigma}y)^2+\mu^2 y^2\right]
	=\sum \omega_n\alpha_n\overline{\alpha_n}
\end{equation}
and
\begin{equation}\label{EPlus}
\hat{\cal E}_+(\gamma)=-\sum_n{\pp\over \omega_n -\cos 2\alpha n}
	\alpha_n\overline{\alpha_n}
\end{equation}
Taking into account (\ref{MuAndJ}) we get
\begin{equation}\label{EandJ}
	{\cal E}(\gamma)=-\epsilon^2\gamma^2 J + \epsilon^2\gamma^2 
	\hat{\cal E}_+(\gamma)
\end{equation}
This is the generating function for the ``left'' Pohlmeyer
charges. In the generating function for the ``right'' charges
we replace $\hat{\cal E}_+(\gamma)$
with $\hat{\cal E}_-(\gamma)$ which is obtained
from $\hat{\cal E}_+(\gamma)$ by replacing 
$n\to -n$. We can also consider the
average of the left and right generating functions:
\begin{equation}
	{\cal E}^{even}(\gamma)=-\epsilon^2\gamma^2 J +
	\epsilon^2\gamma^2 \hat{\cal E}^{even}(\gamma)
\end{equation}
where
\begin{equation}\label{Even}
	\hat{\cal E}^{even}(\gamma)=
	-\sum_n{\pp\omega_n\over \omega_n^2 -(\cos 2\alpha)^2 n^2}
	\alpha_n\overline{\alpha_n}
\end{equation}
\subsection{A relation between ${\cal E}_+$ and ${\cal E}_-$.}
Eq. (\ref{EPlus}) can be also written in the following way:
\begin{eqnarray}\label{WithLambda}
&	\hat{\cal E}_{\pm}(\gamma)=-(1+\gamma^2)
	\sum_n {\mu(\omega\pm n)\over \mu^2+\gamma^2(\omega\pm n)^2}
	\alpha_n\overline{\alpha_n}=\nonumber\\ \\[-9pt]
&	=
	-\sum_n {\mu(\omega\pm n)\over 
	\mu^2+\sin^2\alpha [(\omega\pm n)^2-\mu^2]}\nonumber
\alpha_n\overline{\alpha_n}
\end{eqnarray}
It is usually understood that ${\cal E}(\gamma)$ is 
the generating function of the conserved charges; it is 
defined as a power series in $\gamma$. But
we see from (\ref{WithLambda}) that the expansion in powers of 
$\sin^2\alpha={\gamma^2\over 1+\gamma^2}$ is in fact the expansion
in powers of the expression
$\left[{(\omega\pm n)^2\over\mu^2}-1\right]$ which is small
in the null-surface limit $\mu\to\infty$.
Therefore we can analytically continue ${\cal E}(\gamma)$ to finite
values of $\gamma$. 
It follows from $\omega^2-n^2=\mu^2$ that
\begin{equation}\label{LeftRightSymmetry}
	\hat{\cal E}_{+}(\gamma)=\hat{\cal E}_-(\gamma^{-1})
\end{equation}
This identity follows from the definitions (\ref{BT}), (\ref{GGF}).
Eq. (\ref{BT}) defines the left B\"acklund transformation $Y\to Y'$.
The definition of the right B\"acklund transformation $Y\to Y''$
differs from (\ref{BT}) by $\partial_+\leftrightarrow \partial_-$.
With this definition $Y'(\gamma)=-Y''(-\gamma^{-1})$.
Now (\ref{LeftRightSymmetry}) follows from the formula (\ref{GGF})
for the generating function.

\section{An action variable in the plane wave limit.}
\subsection{Action variable and local conserved charges of the
free massive theory.}
The Hamiltonian of the plane wave theory is:
\begin{equation}
	H={1\over 4\pi} \int d\sigma \sum\limits_{i=1}^4
	\left[p_i^2+q_i^2+
	(\partial_{\sigma}x_i)^2+(\partial_{\sigma}y_i)^2
	+\mu ^2(x_i^2+y_i^2)\right]
\end{equation}
where $p_i(\sigma)=\partial_{\tau}x_i$ and 
$q_i(\sigma)=\partial_{\tau}y_i$ are the momenta conjugate
to $x_i$ and $y_i$ respectively:
\begin{equation}
	\{p_i(\sigma), x_j(\sigma')\}= 2\pi \delta(\sigma-\sigma'),\;\;
	\{q_i(\sigma), y_j(\sigma')\}= 2\pi \delta(\sigma-\sigma')
\end{equation}
Consider the following action variable:
\begin{equation}
	{\cal I}={1\over 4\pi}
	\int d\sigma \left[
	y_i(\sigma) \sqrt{\mu ^2-\partial_{\sigma}^2}\; y_i(\sigma)+
	q_i(\sigma){1\over\sqrt{\mu ^2-\partial_{\sigma}^2}}\; q_i(\sigma) 
	\right]
\end{equation}
This quantity is conserved and generates periodic trajectories. 
In the oscillator language:
\begin{equation}
	{\cal I}=\sum_n \alpha_n\overline{\alpha_n}
\end{equation}
On the other hand the free massive theory has an infinite family
of local conserved charges:
\begin{eqnarray}\label{LocalIntegrals}
&& I_k={1\over 4\pi}\sum\limits_{i=1}^4
\int d\sigma \left[q_i \partial^{2k-2}_{\sigma} q_i - 
y_i \partial^{2k}_{\sigma} y_i
+\mu ^2 y_i\partial_{\sigma}^{2k-2} y_i\right]
=\nonumber \\[5pt]
&&= (-1)^{k+1}\times \sum_n n^{2k-2}\omega_n\; \alpha_n\overline{\alpha_n}
\end{eqnarray}
One can see that ${\cal I}$ is an infinite linear combination of $I_k$:
\begin{eqnarray}\label{PlaneWaveLimit}
	&&	{\cal I}={1\over 4\pi}\sum\limits_{i=1}^4
\int d\sigma \left[ \mu  y_i y_i - 
\sum\limits_{n=0}^{\infty} \mu ^{-2n-1}
{(2n)!\over 2^{2n+1} (n+1)!n!}\; y_i  \partial_{\sigma}^{2n+2} y_i+\right.
\nonumber\\[5pt]
&& 	\left.+\sum\limits_{n=0}^{\infty} \mu ^{-2n-1}
{(2n)!\over 2^{2n} (n!)^2}\; 
q_i  \partial_{\sigma}^{2n} q_i
\right]
=\sum\limits_{n=0}^{\infty}
\mu^{-2n-1} 
{(2n)!\over 2^{2n} (n!)^2} I_{n+1}
\end{eqnarray}
In this section we will show that $I_k$ descend from the
Pohlmeyer charges of the sigma model on $S^5$. We will argue that
Eq.(\ref{PlaneWaveLimit}) provides an expression for
the action variable in terms of the 
local conserved charges (which is valid not only in the plane wave
limit).

\subsection{Integral formula.}
There is an integral formula:
\begin{equation}\label{IntegralFormula}
	{1\over\pi}\int_{-\infty}^{+\infty} 
	{d\gamma\over 1+\gamma^2}\; \hat{\cal E}_+(\gamma)
	={1\over\pi}\int_{-\infty}^{+\infty} 
	{d\gamma\over 1+\gamma^2}\; \hat{\cal E}_-(\gamma)
	=-\sum_n \alpha_n\overline{\alpha_n}
\end{equation}
This means that:
\begin{equation}\label{Integral}
J+\sum_n \alpha_n\overline{\alpha_n}=
	-{1\over\pi}\int_{-\infty}^{+\infty}
	{d\gamma\over \gamma^2(1+\gamma^2)}{\cal E}(\gamma)
\end{equation}
This  is an integral representation for the action variable 
in terms of the conserved charges.
We derived this formula in the plane wave limit, but we conjecture
that it is valid also outside of the plane wave limit.
In other words, the right hand side of (\ref{Integral})
should generate periodic trajectories on the phase space of the
classical string in $AdS_5\times S^5$.

\subsection{Expansion in powers of $n\over \mu$.}
Let us rewrite (\ref{Even}) in the following way:
\begin{equation}\label{WithNu}
	\hat{\cal E}^{even}(\gamma)=
	-\sum_n{\sqrt{1+(n/\mu)^2}\over 1+(n/\mu)^2(\sin 2\alpha)^2}
	\alpha_n\overline{\alpha_n}
\end{equation}
Notice that $(\sin 2\alpha)^2={4\gamma^2\over (1+\gamma^2)^2}$.
Let us define the local conserved charges $\hat{\cal E}_{2k}$ 
as follows:
\begin{equation}
	\hat{\cal E}^{even}(\gamma)=\sum\limits_{k=0}^{\infty}
	\gamma^{2k}\hat{\cal E}_{2k}
\end{equation}
One can verify the following identities:
\begin{eqnarray}
	\hat{\cal E}_0+{1\over 8}\hat{\cal E}_2=
	-\sum (1+o(n^2/\mu^2))\alpha_n\overline{\alpha_n}
\nonumber	\\[5pt]
	\hat{\cal E}_0+{11\over 64}\hat{\cal E}_2
	+{3\over 128}\hat{\cal E}_4=
	-\sum (1+o(n^4/\mu^4))\alpha_n\overline{\alpha_n}
	\nonumber\\[5pt]
	\hat{\cal E}_0+{201\over 1024}\hat{\cal E}_2
	+{11\over 256}\hat{\cal E}_4
	+{5\over 1024}\hat{\cal E}_6=
	-\sum (1+o(n^6/\mu^6))\alpha_n\overline{\alpha_n}
	\nonumber\\[5pt]
	\hat{\cal E}_0+{3461\over 16384}\hat{\cal E}_2
	+{949\over 16384}\hat{\cal E}_4
	+{185\over 16384}\hat{\cal E}_6
	+{35\over 32768}\hat{\cal E}_8=
	-\sum (1+o(n^8/\mu^8))\alpha_n\overline{\alpha_n}
	\nonumber
\end{eqnarray}
These equations agree with the results of \cite{ArutyunovStaudacher,Engquist}
(see the discussion in \cite{Notes}).
In the next order we have:
\begin{eqnarray}
	\hat{\cal E}_0
	+{29011\over 131072}\hat{\cal E}_2
	+{569\over 8192}\hat{\cal E}_4
	+{4661\over 262144}\hat{\cal E}_6
	+{49\over 16384}\hat{\cal E}_8
	+{63\over 262144}\hat{\cal E}_{10}=
	\nonumber\\[5pt]
	=-\sum (1+o(n^{10}/\mu^{10}))
	\alpha_n\overline{\alpha_n}
	\nonumber
\end{eqnarray}

\subsection{An infinite linear combination of local conserved charges.}
Instead of expanding ${\cal E}$ in powers of $\gamma$
let us now expand it in powers of 
$(\sin 2\alpha)^2={4\gamma^2\over (1+\gamma^2)^2}$.
Define ${\cal G}_k$ as follows:
\begin{equation}
	{\cal E}^{even}(\gamma)=\epsilon^2\gamma^2\sum_{k=0}^{\infty}
	\left[ {4\gamma^2\over (1+\gamma^2)^2} \right]^k{\cal G}_k
\end{equation}
Then Eqs. (\ref{LocalIntegrals}) and (\ref{WithNu}) imply
that in the plane wave limit
\begin{equation}\label{PWImproved}
	-{\cal G}_k=\mu^{-1-2k}I_{k+1}+\delta_{k,0}J
\end{equation}
Therefore ${\cal G}_k$ are the ``improved'' currents of 
\cite{ArutyunovStaudacher,Engquist}. Indeed, the density
of the $k$-th improved current is proportional to $|p_S|^{1-2k}$
where $p_S=\partial_{\tau}x_S$ is the push-forward of
$\partial\over\partial\tau$ by the embedding of the
string worldsheet in the $S^5$; it is assumed that
$\tau$ and $\sigma$ are conformal coordinates on the
worldsheet and $\sigma$ is periodic with the period $2\pi$.
The plane wave limit keeps the terms quadratic in $y$.
The typical term is 
$\int d\sigma |p_S(\sigma)|^{1-2k} (\partial_{\sigma}^k Y, 
\partial_{\sigma}^k Y)$. 
In the plane wave limit
$|p_S|\simeq \mu$ and $Y\simeq \epsilon y$.
Therefore in the plane wave limit this quadratic term
will be of the form 
$\epsilon^2\mu^{1-2k} \int d\sigma  (\partial_{\sigma}^k y, 
\partial_{\sigma}^k y)$
which is in agreement with (\ref{PWImproved}) and (\ref{LocalIntegrals}).

Eqs. (\ref{EandJ}) and (\ref{WithNu}) 
imply that the action variable is an infinite sum of ${\cal G}_k$:
\begin{equation}
	J+\sum\limits_{n=-\infty}^{\infty}
	\sum\limits_{i=1}^4\alpha_n^i\overline{\alpha_n^i}=
	-\sum_{k=0}^{\infty}
	{1\over 2^{2k}}{(2k)!\over (k!)^2}\;{\cal G}_k
\end{equation}
We have so far considered only the $S^5$ part of the string worldsheet.
Therefore the Pohlmeyer charges ${\cal G}_k$ which we considered
should be really denoted ${\cal G}_k^{S}$. The index $S$ tells us
that these charges came from the $S^5$ sigma model. One can introduce
in the same way the Pohlmeyer charges ${\cal G}_k^{A}$ in the
$AdS_5$ sigma model. We would get:
\begin{equation}
	E- \sum\limits_{n=-\infty}^{\infty}
	\sum\limits_{i=1}^4\beta^i_n\overline{\beta^i_n} =
	-\sum_{k=0}^{\infty}
	{1\over 2^{2k}}{(2k)!\over (k!)^2}\;{\cal G}_k^A
\end{equation}
Here $\beta_n^i$ are the Fourier modes of $x^i(\tau,\sigma)$.
Arguments of \cite{Anomalous} show that the anomalous dimension
has the following expansion in these charges:
\begin{equation}\label{LogC}
	{1\over 2\pi}\log c^2= 
	\sum\limits_{k=0}^{\infty}
	{1\over 2^{2k}}{(2k)!\over (k!)^2}
	({\cal G}_k^{S}-{\cal G}_k^A)
\end{equation}
We derived this expression from the plane wave limit,
but it should be valid for all classical strings in 
$AdS_5\times S^5$, at least in the perturbative
expansion around the null-surfaces \cite{Notes,KT,Anomalous}.
Eq. (\ref{LogC}) agrees with the usual definition
of the anomalous dimension in the plane wave limit 
as $E-J-\sum\limits_{n=-\infty}^{\infty}\sum\limits_{i=1}^4
(\alpha_n^i\overline{\alpha_n^i}+\beta_n^i\overline{\beta_n^i})$.

\section*{Acknowledgments}
I want to thank I.~Swanson and A.~Tseytlin for interesting discussions,
and M.~Srednicki for asking the question.
This research was supported by the Sherman Fairchild 
Fellowship and in part
by the RFBR Grant No.  03-02-17373 and in part by the 
Russian Grant for the support of the scientific schools
NSh-1999.2003.2.


\begin{thebibliography}{10}
\bibitem{FT02}{S. Frolov, A.A. Tseytlin, 
"Semiclassical quantization of rotating
 superstring in $AdS_5 \times S^5$", JHEP 
{\bf 0206} (2002) 007, hep-th/0204226.}
\bibitem{Tseytlin}{A.A.~Tseytlin, "Semiclassical quantization of superstrings:
$AdS_5\times S^5$ and beyond", Int. J. Mod. Phys. {\bf A18} (2003) 981,
hep-th/0209116.}
\bibitem{Russo}{J.G.~Russo, "Anomalous dimensions in gauge theories from
rotating strings in $AdS_5\times S^5$," JHEP {\bf 0206} (2002) 038, 
hep-th/0205244. }
\bibitem{MinahanZarembo}{J. A. Minahan, K. Zarembo,
	``The Bethe-Ansatz for N=4 Super Yang-Mills,''
	JHEP 0303 (2003) 013, hep-th/0212208.}
\bibitem{FT03}{S. Frolov, A.A. Tseytlin,
	``Multi-spin string solutions in $AdS_5$ x $S^5$,''
	Nucl.Phys. {\bf B668} (2003) 77-110,
	hep-th/0304255.}
\bibitem{FTQ}{S.~Frolov, A.~A.~Tseytlin, ``Quantizing three-spin
	string solution in $AdS_5\times S^5$,''
	JHEP {\bf 0307} 016 (2003), 
	hep-th/0306130.}
\bibitem{Kruczenski}{M.~Kruczenski, 
``Spin chains and string theory'', hep-th/0311203.}
\bibitem{Minahan}{J.~Minahan, 
	``Higher Loops Beyond the SU(2) Sector'',
	hep-th/0405243.}
\bibitem{Notes}{A.~Mikhailov, ``Notes on fast moving strings'',
	hep-th/0409040.}
\bibitem{ArutyunovStaudacher}{ G. Arutyunov, M. Staudacher,
	 ``Matching Higher Conserved Charges for Strings and Spins'',
JHEP 0403 (2004) 004, hep-th/0310182}
\bibitem{Engquist}{J.~Engquist,
	``Higher Conserved Charges and Integrability 
	for Spinning Strings in $AdS_5$ x $S^5$'',
	JHEP 0404 (2004) 002, hep-th/0402092.}
\bibitem{KT}{M. Kruczenski, A. Tseytlin,
	``Semiclassical relativistic strings in 
	$S^5$ and long coherent operators in N=4 SYM theory'',
	hep-th/0406189.}
\bibitem{MandalSuryanarayanaWadia}{G.~Mandal, N.V.~Suryanarayana,
	S.R.~Wadia, 
	``Aspects of Semiclassical
	Strings in AdS$_5$'', Phys.Lett. {\bf B543} (2002) 81,
	hep-th/0206103.}
\bibitem{BPR}{I.~Bena, J.~Polchinski, R.~Roiban,
	``Hidden Symmetries of the $AdS_5 \times S^5$ Superstring'',
	Phys.Rev. {\bf D69} (2004) 046002, hep-th/0305116.}
\bibitem{KMMZ}{V.A.Kazakov, A.Marshakov, J.A.Minahan, K.Zarembo,
	``Classical/quantum integrability in AdS/CFT'',
	JHEP 0405 (2004) 024, hep-th/0402207.}
\bibitem{KazakovZarembo}{ V.A. Kazakov, K. Zarembo,
	``Classical/quantum integrability in non-compact sector of AdS/CFT'',
	hep-th/0410105.}
\bibitem{BKS}{N. Beisert, V. A. Kazakov, K. Sakai,
	``Algebraic Curve for the SO(6) sector of AdS/CFT'',
	hep-th/0410253.}
\bibitem{SchaferNameki}{S.~Schafer-Nameki, 
	``The Algebraic Curve of 1-loop Planar N=4 SYM'',
	Nucl.Phys. {\bf B714} (2005) 3-29, hep-th/0412254.}
\bibitem{Pohlmeyer}{K.~Pohlmeyer, ``Integrable Hamiltonian
	 Systems and Interactions through Quadratic Constraints'',
	 Comm. Math. Phys. {\bf 46} (1976) 207-221.}
\bibitem{BMN}{D.~Berenstein, J.~Maldacena and H.~Nastase,
	"Strings in Flat Space and PP-Waves from ${\cal N}=4$ Super
	Yang-Mills", JHEP {\bf 0204} (2002) 013, hep-th/0202021.}
\bibitem{deVegaLarsenSanchez}{H.J.~de~Vega, A.L.~Larsen, N.~Sanchez,
	``Semi-Classical Quantization of Circular Strings 
	in De Sitter and Anti De Sitter Spacetimes'',
	Phys.Rev. D51 (1995) 6917-6928, hep-th/9410219.}
\bibitem{Anomalous}{A.~Mikhailov, ``Anomalous dimension and local
	charges'', hep-th/0411178.}
\bibitem{Alday}{L. F. Alday, 
	''Non-local charges on $AdS_5 \times S^5$ and PP-waves``,
	JHEP 0312 (2003) 033, hep-th/0310146.}
\bibitem{SwansonMay}{I. Swanson, 
	''On the Integrability of String Theory in $AdS_5 \times S^5$``,
	hep-th/0405172.}
\bibitem{SwansonOct}{I. Swanson,
	''Quantum string integrability and AdS/CFT``,
	hep-th/0410282.}
\bibitem{KRT}{M. Kruczenski, A.V. Ryzhov, A.A. Tseytlin,
	``Large spin limit of $AdS_5$ x $S^5$ string theory 
	and low energy expansion of ferromagnetic spin chains'',
	Nucl.Phys. {\bf B692} (2004) 3-49, hep-th/0403120.} 
\end{thebibliography}
\end{document}